\documentstyle[12pt,epsfig]{article} 
\def\baselinestretch{1.3}
\newcommand{\ba}{\begin{array}}
\newcommand{\ea}{\end{array}}
\newcommand{\bd}{\begin{displaymath}}
\newcommand{\ed}{\end{displaymath}}
\newcommand{\be}{\begin{equation}}
\newcommand{\ee}{\end{equation}}
\newcommand{\bea}{\begin{eqnarray}}
\newcommand{\eea}{\end{eqnarray}}

\def\a{\alpha}

\def\b{\beta}

\def\q2 {q^2}

\def\bt{\begin{table}}
\def\et{\end{table}}

\catcode`@=11 
\def \gsim{\mathrel{\mathpalette\@versim>}}
\def \lsim{\mathrel{\mathpalette\@versim<}}
\def \@versim#1#2{\lower0.4ex\vbox{\baselineskip\z@skip\lineskip\z@skip
     \lineskiplimit\z@\ialign{$\m@th#1\hfil##\hfil$%
     \crcr#2\crcr\sim\crcr}}}
\catcode`@=12 

\begin{document}

\begin{flushright}
{  HRI-P-08-08-002\\
HRI-RECAPP-2008-010}
\end{flushright}

\begin{center}

{\large\bf Non-universal gaugino and scalar masses, hadronically quiet 
trileptons and the Large Hadron Collider}\\[15mm]
Subhaditya Bhattacharya\footnote{E-mail: subha@mri.ernet.in},
AseshKrishna Datta\footnote{E-mail: asesh@mri.ernet.in} 
and Biswarup Mukhopadhyaya\footnote{E-mail: biswarup@mri.ernet.in}\\
{\em Regional Centre for Accelerator-based Particle Physics \\
     Harish-Chandra Research Institute\\
Chhatnag Road, Jhunsi, Allahabad - 211 019, India}
\\[20mm] 
\end{center}

\begin{abstract} 
We investigate the parameter space of the minimal supersymmetric 
Standard Model (MSSM) where the gluino and squark masses are much above 
1 TeV but the remaining part of the sparticle spectrum is accessible to the
Large Hadron Collider at CERN. After pointing out that such a scenario
may constitute an important benchmark of gaugino/scalar non-universality, 
we find that hadronically quiet trileptons are rather useful signals
for it. Regions of the parameter space,
where the signal is likely to be appreciable, are identified through 
a detailed scan. The advantage of hadronically quiet trileptons over
other types of signals is demonstrated.
   
\end{abstract}

\vskip 1 true cm

\newpage
\setcounter{footnote}{0}

\def\baselinestretch{1.5}

\section{Introduction}

The search for physics beyond the Standard Model (SM) of 
elementary particles enters into an exciting phase as
the Large Hadron Collider (LHC) at CERN takes off. 
A  leading candidate among new physics
options is supersymmetry (SUSY) that can survive down to the
TeV scale \cite{Book,Sally,Martin,Martin1}. 
Hence, the search for SUSY in the context of the LHC has become
one of the major areas of recent studies in particle physics 
\cite{Gl,Bourjaily:2005ja}.

The minimal SUSY Standard Model (MSSM), comprising all the SM particles
and their superpartners, has a large number ($\simeq 100$) of free 
parameters. Some organizing principle is therefore sought to
simplify the picture, the popular paradigm being an approach where
all of the low-scale parameters evolve out of a select few 
at a high scale, possibly related to a Grand Unified Theory (GUT) 
\cite{GUT1}. A frequently studied possibility in this context is 
a scenario based on supergravity (SUGRA) where SUSY breaking gaugino 
and scalar masses, get mutually related in a model dependent manner 
at low energy \cite{Book,Martin,mSUGRA1,mSUGRA2}. However, 
such relationship is contingent upon additional 
features such as the absence of physics at intermediate scales.

The predicted signals of SUSY at the LHC depend largely
on the production of strongly interacting superparticles, namely, the
squarks and gluinos, via the annihilation of quarks, antiquarks and 
gluons. Their subsequent decays culminate into the lightest SUSY
particle (LSP)-- the lightest neutralino in most scenarios-- which
is stable and a dark matter candidate when R-parity is conserved. The 
resulting signals of SUSY consist in a large amount of missing
transverse energy ($E_{T}\!\!\!\!\!\!/$ ), together with hard central 
jets and leptons of various multiplicities \cite{CDF,baer11}. Of course, 
hadronically quiet events, such as trileptons arising from the direct 
production of charginos and neutralinos have also been studied as useful 
supporting signals which can help in probing the non-strongly interacting 
sector of the theory \cite{hq3l}. Still, the final states arising from 
squark/gluino cascades have overwhelming importance in general, for the 
sheer level of copiousness that they associate.  

How about a situation where the squarks and gluinos 
are so heavy (say, $\gsim$ 5 TeV) that
their production rate is too low to support the cascades?
In such a case, most of the signals that depend on strongly 
interacting superparticles will not be easy to see at the LHC. Keeping
such a situation in mind, it is important in one's preparation for the
LHC to check how the `hadronically quiet events' fare. One needs to know
exactly over which ranges in the parameter space of SUSY, for
example, {\em the hadronically quiet trilepton events can act as the
harbinger of new physics, not as a supplementary search channel but
as the main one}. The present work is aimed at answering such a
question, by making an elaborate survey of the SUSY parameter space,
especially in terms of $M_1$ and $M_2$, the U(1) and SU(2) gaugino
masses which dictate the rate of the hadronically trilepon rates.
In other words, although hadronically quiet trileptons in the context
of the LHC have been already studied, here we wish to make the study
specifically focused on cases where squarks and gluinos tend to decouple.

Hadronically quiet trileptons occur mostly from the production 
$pp \longrightarrow \chi_2^0 \chi_1^{\pm}$, where $\chi_2^0$
is the second lightest neutralino and  $\chi_1^{\pm}$ is the
lighter chargino. The hadronically quiet trilepton events have the  
best chance when the squarks are very heavy compared to the sleptons and 
decays of charginos and neutralinos to on-shell sleptons and leptons are 
allowed. From this point of view, the decoupled
nature of squarks favours the trilepton final states. On the other hand, 
they have less of a chance when the decay modes 
$\chi_2^0 \longrightarrow  \chi_1^0 h^0$ or
$\chi_2^0 \longrightarrow  \chi_1^0 Z$  have substantial 
branching ratios.

An exhaustive investigation of the SUSY parameter space in this light has to 
go beyond universality of gauginos and scalars at high scale.
We outline some ways of theoretically motivating non-universality 
in the next section. However, we wish to re-iterate that, in the absence
of any concrete knowledge of high scale physics as well as whether a 
`grand desert' exists, it is really important in the context of collider 
studies to go beyond specific theoretical schemes. {\em We rather 
propose to establish a new benchmark of non-universal SUSY breaking 
masses, distinguished by the suppression of final states arising from 
strong production}. The remaining part of the paper is an exercise in this 
direction and therefore in our collider studies we treat the strong versus 
electroweak gaugino and scalar masses essentially as phenomenological inputs.

In section 2 we first outline some standard GUT-based schemes of
achieving non-universal SUSY breaking masses. Then it is shown that
the situation with heavy squarks and gluinos may require one to go
beyond such schemes, and establish our benchmark based on this
consideration. A detailed discussion of the hadronically quiet trilpton
signal, and the main backgrounds, is presented in section 3.
In section 4 we present numerical predictions for leptonic final
states of various multiplicity, with accompanying hard jets.
We conclude in section 5.

\section{Non-universality and hadronically quiet signals} 

The kind of spectrum that we use for our study can be motivated from
the non-universality of gaugino and scalar masses at  high scale.

As is well known, universality of gaugino masses at high scale is not a 
necessity even in GUT-based scenarios. A number of nonuniversal ratios 
among $M_{1,2,3}$ can arise, say, in $SU(5)$ and $SO(10)$ scenarios, with 
general gauge kinetic functions

\bea
f_{\a \b}(\Phi^{j})= f_{0}(\Phi^{S})\delta_{\a
  \b}+\sum_{N}\xi_{N}(\Phi^s)
\frac{{\Phi^{N}}_{\a \b}}{M}+ {\mathcal{O}}(\frac{\Phi^N}{M})^2
\eea
\noindent
where $f_0$ and $\xi^N$ are functions of chiral singlet superfields, and
$M$ is the reduced Planck mass$=M_{Pl}/\sqrt{8\pi}$. Here $\Phi^{S}$
and $\Phi^{N}$ are Higgs multiplets that are, respectively, singlets
and non-singlets under the GUT group. Different non-singlet representations
leading to the breaking of the GUT group, arising from symmetric products of
the adjoint representations, lead to different ratios among
the high-scale values of the three gaugino masses 
\cite{Ellis, nonugmpheno2, Subho1}. However, in neither of the cases 
pertaining to the two GUT groups mentioned above can one have 
$M_3 >> M_{1,2}$ at the electroweak scale (see Table 2). One cannot achieve 
the above hierarchy by breaking the GUT group via linear combinations of 
various non-singlet representations unless there is strong cancellation among 
various contributing multiplets.
 
Nonetheless, as has been already mentioned, a hierarchy of the gluino and 
electroweak gaugino masses can arise from hitherto unknown effects, such as 
the presence of intermediate scale(s) as well as the evolution
between the Planck and the GUT scales. 
\vspace {0.5 cm}
\noindent
\begin{center}
\begin{tabular}{|c|c|c|c|c|c|c|c|}
\hline
 $m_{\tilde{\ell}}$ in GeV & $(M_1,M_2)$ in GeV  & $OSD$ & $SSD$ & 
$3\ell ~+~jets$ & $\ge 3~jets$ & $3\ell$ \\
\hline 
 200 & (150,300) & 1.25 & 0.04 & 0.11 & 2.82 & 5.99 \\
\hline 
 300 & (232,350) & 0.55 & 0.07 & 0.10 & 1.79 & 2.39 \\
\hline
400 & (179,200) & 0.24 & 0.07 & 0.01 & 3.37 & 0.11 \\
\hline
\end{tabular}
\end{center}
\vspace {0.5 cm}
{Table 1: Different final state rates (fb) with cuts at the LHC with 
$M_3=m_{\tilde g}$= 5 TeV, $m_{\tilde q}$= 5 TeV, $\mu$= 1 TeV, 
$m_A$= 500 GeV, $A$ = 0, $\tan \beta$= 10, where $\mu$, $A$ and 
$\tan \beta$ are respectively the Higgsino mass parameter, the trilinear soft 
SUSY breaking parameter and the ratio of the vacuum expectation values of the 
two Higgs doublets. All the parameters are at the electroweak scale with 
appropriate mixing in the third family. $\ell$ stands for electrons and
muons. {\tt CTEQ5L} PDFset used with $\mu_{F}=\mu_{R}=~ \sqrt{\hat s}$.}\\

In the scalar sector, while certain SUSY-GUT effects like $SO(10)$
$D$-terms can lead to non-universality of mass parameters at high scale 
\cite {so101, so102, Datta:1999uh, subho2},
it is generally difficult to accommodate squarks much heavier than
sleptons in such a framework. One cannot however rule out, for
example, additional $U(1)$ symmetries under which the squarks and 
sleptons have widely disparate charges, and which breaks to make
the squarks much heavier than sleptons via $D$-terms. In addition,
if a large hierarchy exists in the gaugino sector, making the
$SU(3)$ gaugino mass much higher than those of the $SU(2)$ and $U(1)$ 
gauginos at high scale, then even a universal scalar mass scenario
can make the squarks much heavier at the electroweak scale, through the
large gluino contribution in the process of running.

In the rest of our study we take the
low-energy spectrum as a phenomenological input, and look at regions
where large squark and gluino masses prevent strong processes from
contributing significantly to SUSY signals at the LHC. We wish to see SUSY 
signals when, in the above situation, the sleptons and electroweak gauginos 
are well within the reach of the machine.

We show in Table 1 three sample points in situations of the above type. 
These points are consistent with the cold dark matter relic density indicated 
by the WMAP results ($0.91<\Omega_{CDM}h^2<0.128$ within 3$\sigma$ limit) 
\cite {DM}. The relic density for these points have been computed using the 
{\tt SLHA} output of the low-energy SUSY spectra from {\tt Suspect} v2.3 
\cite{SuSpect} and feeding it to the code {\tt micrOMEGAs} v2.0 \cite{DM1}. 
Corresponding to these points, rates are presented for opposite-sign 
dileptons (OSD), same-sign dileptons (SSD) and trilepton final states 
($3\ell ~+~jets$) each associated with hard central jets, as also for the
inclusive jets ($\ge 3~jets$). Lastly, the hadronically quiet trilepton 
($3\ell$) rate is presented, each case being characterized by missing $E_T$. 
Acceptance cuts as specified in our earlier works \cite{Subho1,subho2} 
have been used in computing these rates.
It can be seen that all these rates are suppressed in this region of the 
parameter space. Compared to them, the rate for hadronically quiet trileptons
arising from purely electroweak processes turns out to be higher,
though they are still somewhat small in the absolute sense. The points chosen 
in Table 1 are samples, where the statistical significance of the signals over
backgrounds is not as much the issue as the relative strengths of the 
hadronically quiet trileptons vis-a-vis other signals. We show after
a detailed scan of the parameter space that the hadronically
quiet trilepton signal, largely the result of $\chi^{\pm}_1\chi^0_2$ 
production, is still significant over a noticeable region of the parameter 
space.

It is in general seen that the signals are appreciable, and 
simultaneously the WMAP bound can be satisfied with relative ease,
if the slepton mass in on the low side ($\lsim$ 300
GeV). For $m_{\tilde {\ell}}$ = 200 GeV, the WMAP-allowed region
spans over $M_1$ in the range between 103 GeV and $\gsim$ 175
GeV, while $M_2$ varies in the range 120 - 300 GeV.
For larger slepton masses, the allowed band shifts 
to larger valus of $M_1$ (approximately 170 - 235 GeV for
a slepton mass of 300 GeV) for the same $M_2$.
The allowed band includes regions of lower $M_1$ and $M_2$ for
lower values of $\mu$ where, however, the hadronically quiet signals
become more intractable, as the enhanced Higgsino components in $\chi^{\pm}_1$
and $\chi^0_2$ reduce their couplings to leptons of the first two families.

While the sample points shown in Table 1 are fully consistent with
the WMAP constraints, and serve to illustrate the efficacy of the
hadronically quiet trilepton channel, we feel that a scan over a large
region of the parameter space should be made in an analysis pertaining
to the LHC. In this spirit, we have calculated the signal rates in the
entire region over the $M_1 - M_2$ space allowed by terrestrial
experiments, with various values of the slepton mass, assuming that 
the squark and gluino masses are 5 TeV (where they contribute little to
the cascades). Apart from the values of $M_1$, $M_2$ and the slepton mass,
all the other SUSY parameters are fixed at values used in Table 1 for
most of our analysis. Variation with squark/gluino mass and $\tan \beta$
are shown only at the end of the next section, to demonstrate how
they affect the predictions.

We indicate in Table 2 some sample high scale parameters that generate a 
representative SUSY spectrum in our benchmark scenario, running two loop 
renormalisation group equation (RGE) with radiative corrections to all
squark and gaugino masses in {\tt Suspect} v2.3. It has been 
obtained by using the pMSSM option of the code. It is demonstrated that 
non-universality in the gaugino sector can be responsible for the kind of
spectrum phenomenologically adopted by us. It should be noted that
the non-universality of $M_3$ with $M_{1,2}$ required here, 
can be produced within the ambit of familiar SUSY-GUT, but with a strong 
cancellation between different contributing non-singlet representations, 
as mentioned earlier.

\noindent
\begin{center}
\begin{tabular}{|c|c|c|c|c|c|c|} 
\hline
{GUT-Scale input} & $M_1$ & $M_2$ & $M_3$ & $m_{0\tilde {\ell}}$
& $m_{0\tilde q}$ & $sgn(\mu)$ \\
\hline
 & 300 & 300 & 2400 & 300 & 300 & +ve\\
\hline 
{Low-Scale Output} & $M_1$ & $M_2$ & $m_{\tilde g}$ & $m_{\tilde {\ell}}$&
 $m_{\tilde q}$ & $\mu$ \\ 
\hline
 & 113.8 & 194.0 & 4961.4 & $\simeq$300 &$\simeq$4200 & 2630\\
\hline
\end{tabular}
\end{center}
\vspace {0.5 cm}
{Table 2: Spectrum (in GeV) generated with {\tt Suspect} v2.3 by having 
high scale gaugino mass non-universality. $\tan\beta$= 10, $A_0$= 0.  
Radiative electroweak symmetry breaking is ensured. High scale Higgs mass 
parameters ${m_{H_u}}^2$ and ${m_{H_d}}^2$ are kept degenerate with universal 
scalar masses ($m_{0\tilde{\ell}}= m_{0\tilde{q}}$) at the same scale.}\\

\section{Signal and backgrounds: hadronically quiet trileptons}

We have used the event generator {\tt Pythia} v6.4.16 \cite{PYTHIA} for the 
generation of low-energy SUSY spectra. The consistency of
parameter combinations under investigation have been checked with the
the programme {\tt Suspect} v2.3, where all the low-energy constraints from
$b\longrightarrow s\gamma$, muon anomalous magnetic moment etc.  are
taken into account \cite{constraints}. The Higgsino mass parameter 
$\mu$ is used as a free parameter in the numerical study.

{\tt Pythia} v6.4.16 has also been used for the 
simulation of $pp$ collision with the centre-of-mass energy of 
14 TeV, with hadronization effects turned on. 
We have used {\tt CTEQ5L} \cite{CTEQ}  parton distribution functions,
the QCD renormalization and factorization scales ($\mu_{R},\mu_{F}$) being
both set at the subprocess centre-of-mass energy  $\sqrt{\hat{s}}$.
As we shall show later, the overall conclusions are rather
insensitive to the choice of scales.

All possible SUSY processes and decay chains consistent 
with conserved $R$-parity have been kept open. 
We have switched on initial and final state radiations 
(ISR and FSR respectively) with the functions built within {\tt Pythia} 
v6.4.16, but otherwise confined ourselves to the lowest order matrix elements 
for the signal. The effect of multiple interactions has been neglected.

Jets are formed in {\tt Pythia} using {\tt PYCELL} jet formation 
criteria with $|{\eta}_{jet}| ~\le ~5$ in the calorimeter, 
$N_{\eta_{bin}}=100$ and $N_{\phi_{bin}}=64$. 
For a partonic jet to be considered as a jet initiator, $E_{T}> 2$ GeV 
is required, while a cluster of partonic jets is branded as a hadron-jet
when $\sum_{parton} E_{T}^{jet}$ is more than 20 GeV.
The maximum $\bigtriangleup R$ from the initiator jet is taken to be 0.4. 
We have cross-checked the hard scattering cross-sections 
of various production processes with {\tt CalcHEP} \cite{CalcHEP}.
All the final states with jets at the parton level 
have been checked against the results available in \cite{Ash}. 
The calculation of hadronically quiet trilepton 
rates have been checked against other standard works, in the 
appropriate limits \cite{hq3l}.

While the minimum $E_T$ or trigger for jet formation is
20 GeV, hadronically quiet trilepton events (with $\ell=e,\mu$) have been
defined following our earlier work \cite{subho2}. With this definition, 
the absence of any accompanying central jet ($|{\eta}_{jet}| ~\le ~2.5$) with 
$E_{T}^{jet} ~\geq ~100$ GeV qualifies the event 
as hadronically quiet. This avoids unnecessary vetoing of trilepton 
events along with jets originating from ISR/FSR, underlying events 
and pile-up effects. Strong cascades with events leading
to relatively soft jets also add to the signal.

The background to the proposed signal can come from a number of processes 
including  $WZ/Z^{*}/\gamma^{*}$, $t\bar t$ as well as heavy flavours.
The $WZ^{*}/W\gamma^{*}$ and heavy flavour (mostly $b$) channels
are brought under control with a large missing-$E_T$ cut \cite{Berger2}. 
Furthermore, we have demanded the three leptons to be isolated, 
according to the criteria listed below.
In addition, at least one pair of opposite charged leptons
(electrons/muons) have to be of the same flavour. This finally leaves
us with $t\bar t$ and $WZ$ production. Of the latter channel,
whatever survives the missing-$E_T$ cut is suppressed
by imposing an invariant mass cut on opposite-sign, same flavour 
dileptons. Thus it is the  $t\bar t$ channel that really constitutes
the irreducible background, mostly due to the overwhelmingly 
large rate of top-quark pair production at the LHC.

We have generated all dominant SM events  
in {\tt Pythia} for the same final states, using the same  
renormalization/factorization scale, parton distributions and cuts. The $WZ$
and $t {\bar t}$ channels are dominant among the backgrounds.
While the former is effectively suppressed through an invariant 
mass cut on the same flavour, opposite-sign lepton pairs, the 
$t {\bar t}$ background is of an irreducible nature, since,
with the huge production cross-section at the LHC, jets that
do not satisfy either the trigger or our imposed cuts can masquerade
as hadronically quiet events. An enhancement of statistical significance 
of the signal over such backgrounds is attempted with the help
of the missing $E_T$ cut. As we shall see in the
numerical results, a higher degree of significance is 
expected when the mass differences between the $\chi^0_1$ and
each of the $\chi^0_2$ and the $\chi^\pm_1$ are on the higher side,
thus allowing a harder $p_{T}$ spectrum for the leptons. The other backgrounds,
namely, the ones from virtual $Z$/photons, are found to be under
control after imposing the cuts, which are as follows \cite{ATLAS}:

\begin{itemize}
\item  Missing transverse energy $E_{T}\!\!\!\!\!\!/$~ $\geq ~100$ GeV

\item $p_{T}^\ell ~\ge ~20$ GeV and $|{{\eta}}_{\ell}| ~\le ~2.5$
 
\item Lepton isolation, such that lepton-lepton separation
 ${\bigtriangleup R}_{\ell\ell}~ \geq 0.2$, lepton-jet separation 
 ${\bigtriangleup R}_{{\ell}j}~ \geq 0.4$. The ${E_{T}}$ deposit 
due to jet activity around a lepton ${E_{T}}$ within a cone of
$\bigtriangleup R~ \leq 0.2$ of the lepton axis should be $< 10$ GeV

\item No jet with  
$E_{T}^{jet} ~\geq ~100$ GeV and $|{\eta}_{jet}| ~\le ~2.5$ (Vetoing central
  hard jets)

\item Invariant mass of any same flavour, opposite
sign lepton pair with $|m_{Z}-M_{\ell_{+}\ell_{-}}| ~\geq 10$ GeV 

\end {itemize}

\noindent
where  $\bigtriangleup R = \sqrt {{\bigtriangleup \eta}^2
+ {\bigtriangleup \phi}^2}$ 
is the so-called isolation parameter which is the separation in 
the pseudo-rapidity and the azimuthal angle plane.

\noindent
\begin{center}
\begin{tabular}{|c|c|c|c|}
\hline
 Cuts & $\sigma_{t\bar t~\longrightarrow~3\ell}$
 &$\sigma_{WZ~\longrightarrow~3\ell}$ 
& $\sigma_{3\ell(total)}$\\ 
\hline
$p_T,\eta~ cut~ (on~ \ell, jets)$ & 2.428 & 0.130 & 2.557\\
\hline 
$+lepton~isolation$ & 0.473 & 0.031 & 0.504\\ 
\hline
$+E_{T}\!\!\!\!/~ cut$ 
& 0.267 & 0.010 & 0.277\\
\hline
 $+ invariant~ mass~ cut$ & 0.129 & 0.008 & 0.137\\
\hline
\end {tabular}
\end {center}

{Table 3 : Cross-sections (pb) for leading sources of SM background after 
successive application of different cuts, $m_t=171.4$ GeV. {\tt CTEQ5L} PDFset
used with $\mu_{F}=\mu_{R}=~ \sqrt{\hat s}$. 
$\sigma_{t\bar t~\longrightarrow~3\ell}$ is presented after multiplying 
by appropriate $K$-factor (2.04).}\\

\begin{figure}[htbp]
\begin{center}
\vspace*{-2.0cm}
\centerline{\psfig{file=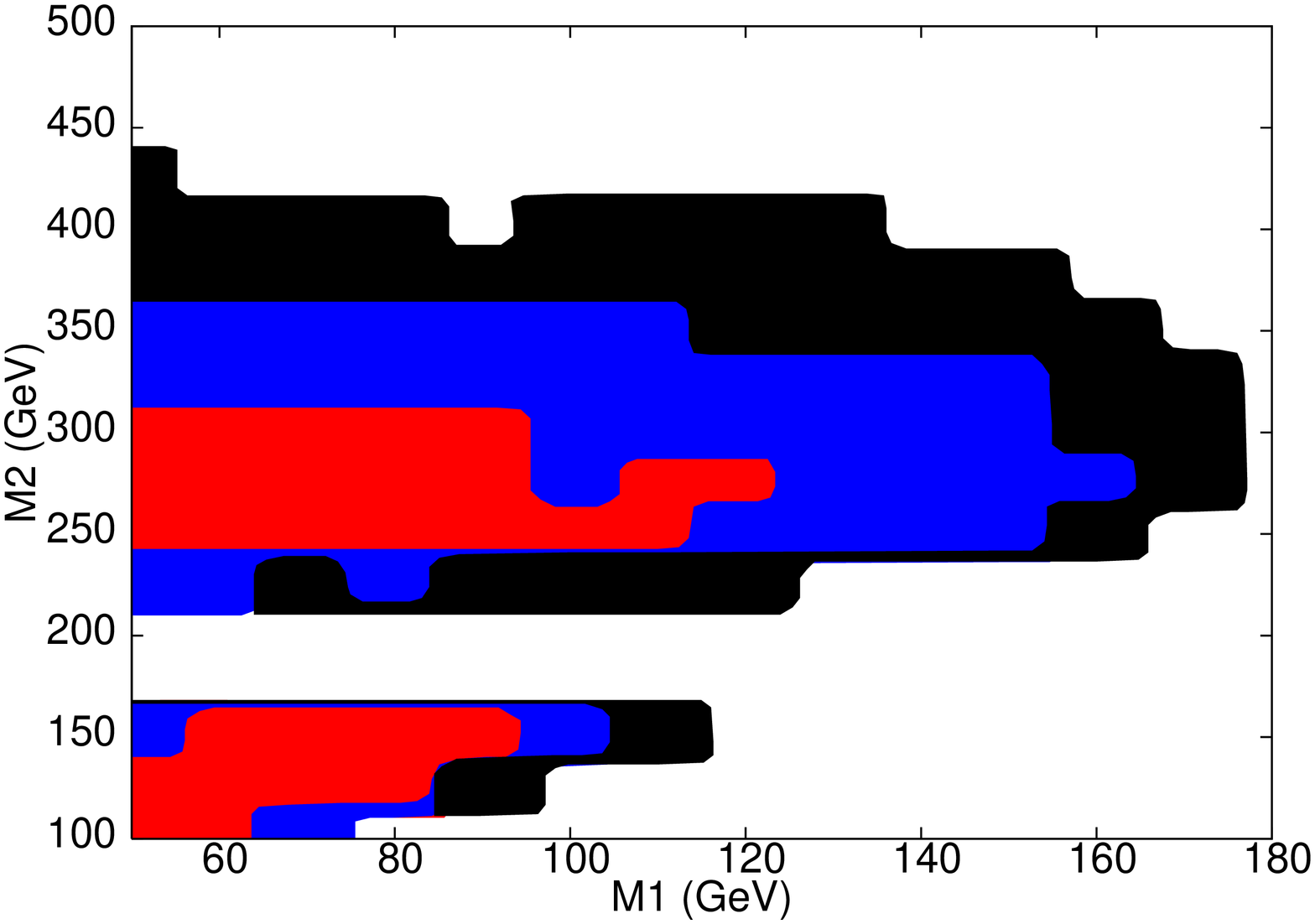,width=6.5 cm,height=7.5cm,angle=-90}
\hskip 20pt \psfig{file=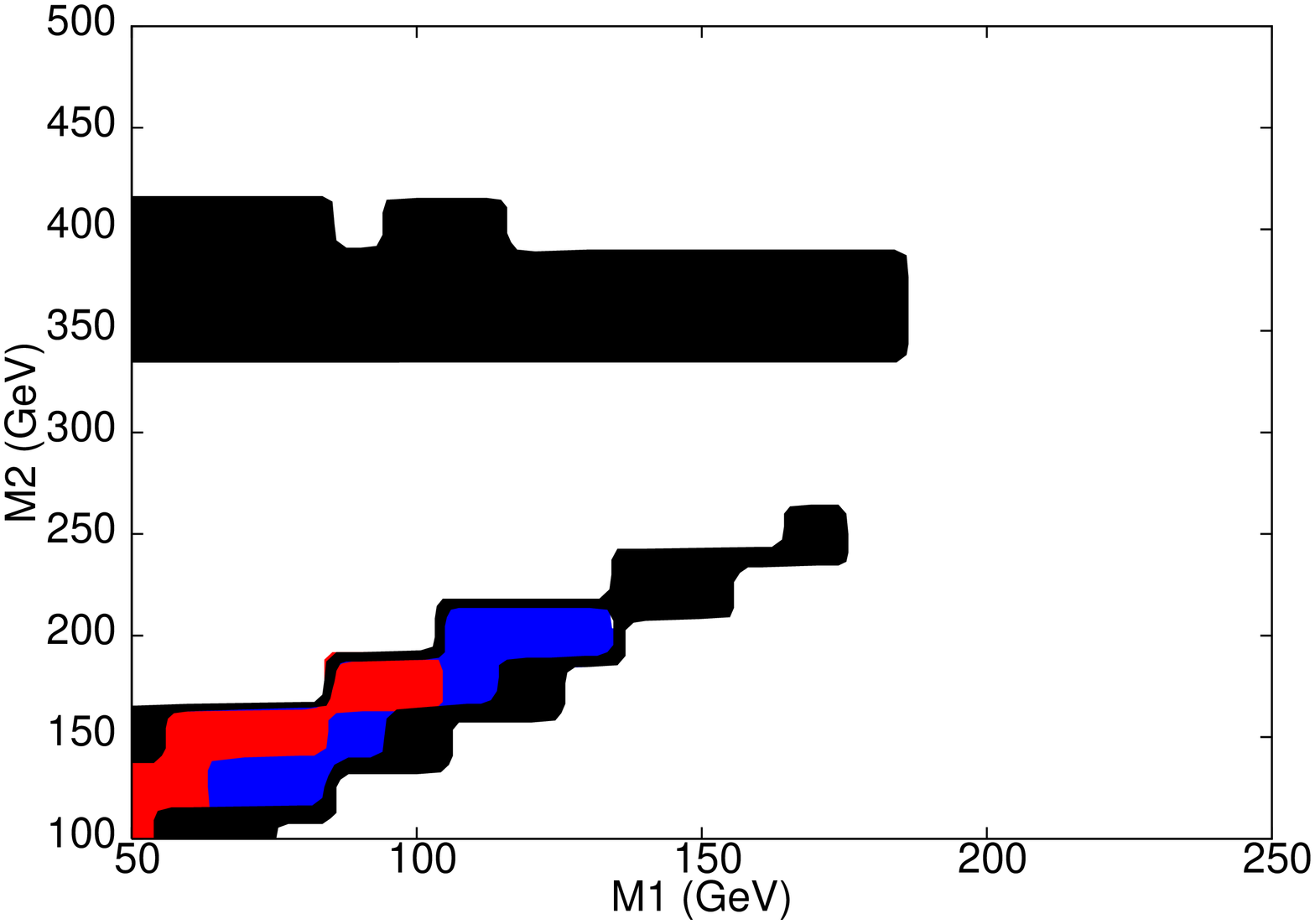,width=6.5 cm,height=7.5cm,angle=-90}}
\vskip 10pt
\centerline{\psfig{file=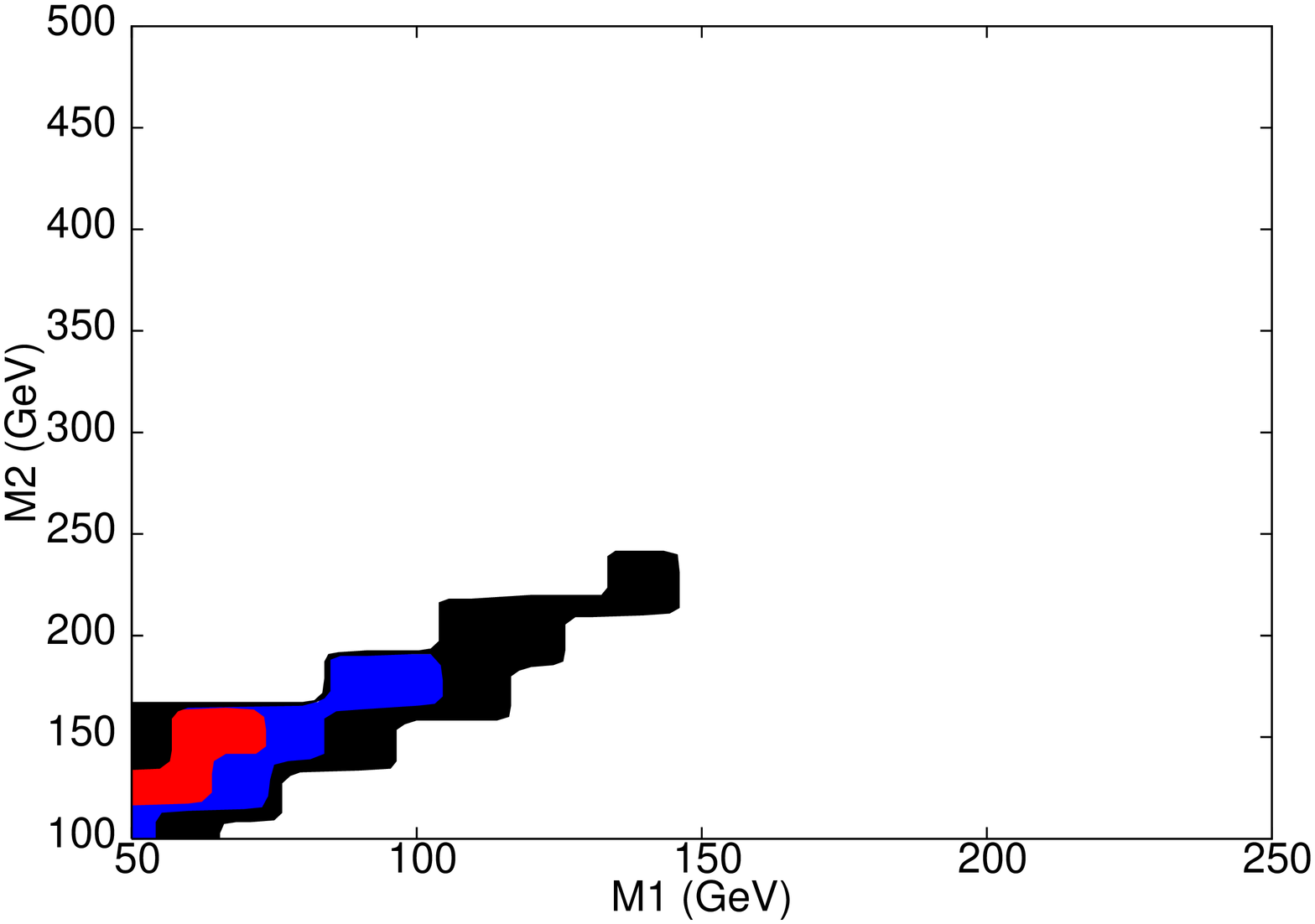,width=6.5 cm,height=7.5cm,angle=-90}}
\caption{Significance contours for hadronically quiet trilepton events, 
for an integrated luminosity of 100 fb$^{-1}$ in $M_1-M_2$ plane. 
{\it Colour Code: {\tt Black:} 2$\le\sigma<$3, {\tt Blue:} 3$\le \sigma<$5, 
{\tt Red:} $\sigma$$\ge$ 5 (in black-and-white print: 
{\tt Black:} 2$\le\sigma<$3, 
{\tt Grey:} 3$\le \sigma<$5, {\tt Light Grey:} $\sigma$$\ge$ 5}). 
Top left: Slepton mass = 200 GeV, Top right: Slepton mass = 300 GeV, 
Bottom: Slepton mass = 400 GeV. {\tt CTEQ5L} PDFset
used with $\mu_{F}=\mu_{R}=~ \sqrt{\hat s}$.} 
\end{center}
\end{figure}

The cross-sections for the backgrounds from the dominant sources, 
subjected to the cuts that are used in the signal-analysis, are 
presented in Table 3. The effectiveness of cuts at successive levels have
been shown.

The numbers of signal and background events have been calculated for
an integrated luminosity of 100 fb$^{-1}$.
The significance is obtained in the 
Gaussian limit, using  $\sigma=S/{\sqrt {B}}$ where $S$ and $B$ 
denote the number of signal and the background events respectively.

In Figure 1 we plot the significance of hadronically quiet trileptons in the 
$M_{1} - M_{2}$ plane, for three different slepton masses 
which are all kept to be degenerate at 200 GeV, 300 GeV and 400 GeV.
Of course, the lighter stau is somewhat lighter than the other sleptons,
and we truncate the value of $M_1$ accordingly in each
plot, so as to disallow a scenario with stau as the 
lightest SUSY particle (LSP). 
In each case, the gluino and squark masses are kept at 5 TeV, with
$\mu$ = 1 TeV and $\tan \beta$= 10. While regions with
less than 2$\sigma$ have not been marked,
regions marked in {\tt red} correspond to significance more than 5, 
{\tt blue}, to significance  in the 3-5$\sigma$ range and {\tt black}, 
to the 2-3$\sigma$ range, while in black-and-white print 
{\tt light grey}, {\tt grey} and {\tt black} implies above 
significance respectively.

For a slepton mass of  200 GeV (the top left plot), 
there is a large region of parameter space for $M_{1}$ between 
50 GeV and 125 GeV and $M_{2}$ between 240 GeV and 300 GeV with
significance more than 5$\sigma$. There also exists a small region 
at the bottom left portion of the graph for $M_{1}$ between 50 GeV and 90 GeV 
and for low $M_{2}$ (between 100 GeV and 140 GeV) which has significance 
more than 5$\sigma$. The regions of significance between 3-5$\sigma$ and 
2-3$\sigma$ lie around the region of $\sigma \ge $5. The statistical 
significance in various regions can be explained
by remembering that the rate of $\chi^{\pm}_1 \chi^0_2$ production
is large for smaller chargino and neutralino masses, thus giving
higher overall rates. At the same time, there is a complementary trend of
a larger number of events surviving the hardness cut once one has
larger $M_2$, thus creating a rather large region in the parameter
space with higher significance of the signal. In addition, there is
a dynamical effect \cite{Baer:1992dc}, namely, the destructive interference
between the $Z$-and slepton-mediated diagrams in $\chi^0_2$ decays,
when on-shell sleptons are not produced. The observed pattern of 
significance contours is a consequence of such effects as well.  

\begin{figure}[htbp]
\begin{center}
\vspace*{-0.0cm}
\centerline{\psfig{file=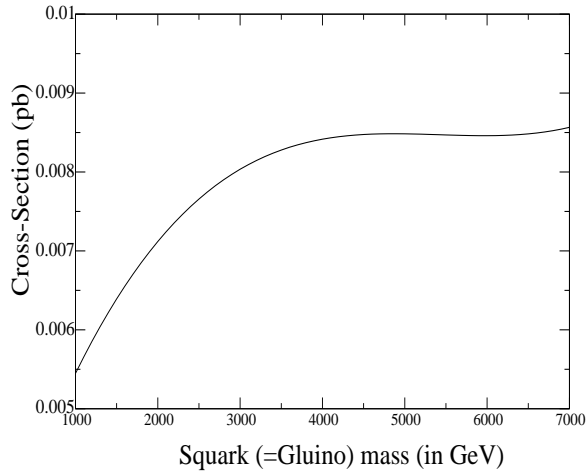,width=7.5 cm,height=8.50cm,angle=-90}}
\caption{Variation of rates (pb) with cuts for hadronically quiet 
trilepton events with degenerate squark-gluino mass. Other relevant parameters
are at the following values: $m_{\tilde \ell}$ = 200 GeV, $M_{1}$ = 100 GeV 
and $M_{2}$ = 300 GeV, $\mu$ = 1 TeV, and $\tan \beta$ = 10. 
{\tt CTEQ5L} PDFset used with $\mu_{F}=\mu_{R}=~ \sqrt{\hat s}$.}
\end{center}
\end{figure}

For higher slepton masses, namely, $m_{\tilde \ell}$= 300 or 400 GeV, 
the region of the parameter space depicting $\sigma \ge $5 for 
$m_{\tilde \ell}$= 200 GeV shrinks. Only the small region at the bottom left 
corner of the graph shows $\sigma \ge $5, although it also shrinks to a 
considerable extent compared to the case of $m_{\tilde \ell}$= 200 GeV.

However, for the case of $m_{\tilde \ell}$= 300 GeV, although 
the 3-5$\sigma$ region is absent in the upper segment, 
the region of 2-3$\sigma$ extends upto 
$M_{1} =$ 180 GeV, and for $M_{2}$ slightly on the higher side 
(340 GeV to 400 GeV). This is because, with the degenerate slepton masses 
going up, the allowed region with neutralino LSP is larger, 
and at the same time the leptons in the final state tend to be harder.
The regions with $\sigma \ge $5 correspond to
regions with  very low $M_{2}$ (110 GeV to 160 GeV) for a slepton
mass of 400 GeV. The erstwhile regions of high significance for larger 
values of $M_2$ are gone for heavier sleptons. In such cases, as has
been mentioned earlier, the $\chi^0_1 h$ and  $\chi^0_1 Z$ channels
tend to dominate  in the decays of $\chi^0_2$, thus reducing the 
significance of the trilepton signals.

We have also checked the dependence of our predictions on
the QCD renormalization/factorization scales by setting,
for instance, both the scales
at the average mass of the final state particles in the hard scattering. 
While this affects both signal and background rates, the significance 
contours remain very similar to the corresponding case with the scale 
set at the subprocess centre-of-mass energy. This shows the robustness of
the expected significance levels.

\begin{figure}[htbp]
\begin{center}
\vspace*{-0.0cm}
\centerline{\psfig{file=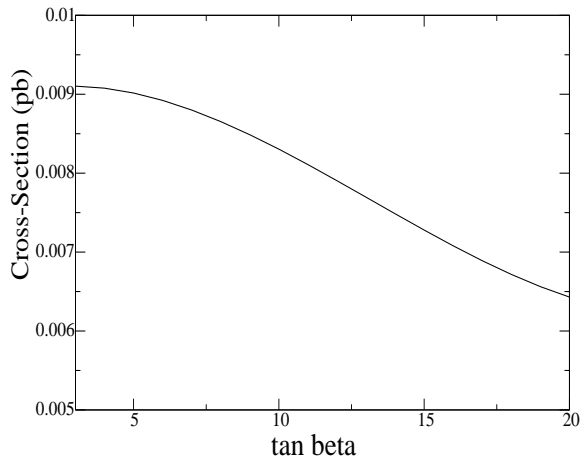,width=7.5 cm,height=8.50cm,angle=-90}}
\caption{Variation of rates (pb) with cuts for hadronically quiet 
trilepton events with $\tan \beta$. Other relevant parameters are at the
following values: $m_{\tilde g}= m_{\tilde q}$ = 5 TeV, $m_{\tilde \ell}$ = 200 GeV, $M_{1}$ = 100 GeV, $M_{2}$ = 300 GeV, 
and $\mu$ = 1 TeV. {\tt CTEQ5L} PDFset used with 
$\mu_{F}=\mu_{R}=~ \sqrt{\hat s}$.}
\end{center}
\end{figure}

We also plot in Figure 2 the variation in rates for hadronically 
quiet trilepton ($3\ell$) events with $m_{\tilde g}= m_{\tilde q}$
varying from 1 to 7 TeV for $m_{\tilde \ell}$ = 200 GeV, 
$M_{1}$ = 100 GeV, $M_{2}$ = 300 GeV, $\mu$ = 1 TeV with $\tan \beta$ = 10.
The rate for hadronically quiet trileptons increases gradually with
the coloured sparticle mass going up, due to the interference between 
the $s$-and squark-mediated $t$-channel diagrams. 
The effect dwindles as the squark and gluino mass reaches 3 TeV, and a plateau 
is clear from about 5 TeV onwards.

\begin{figure}[htbp]
\begin{center}
\vspace*{-0.0cm}
\centerline{\psfig{file=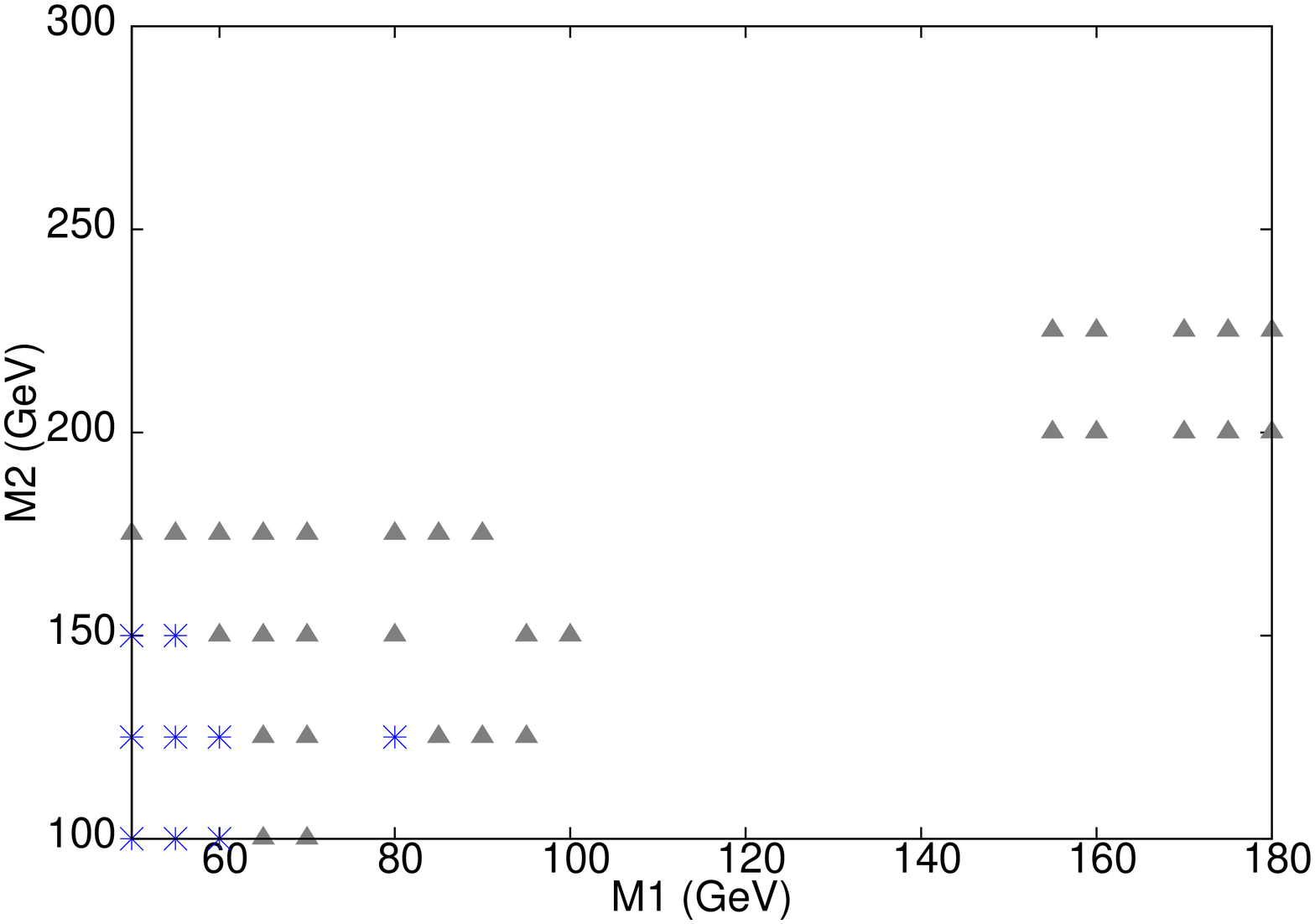,width=6.5 cm,height=7.5cm,angle=-90}
\hskip 20pt \psfig{file=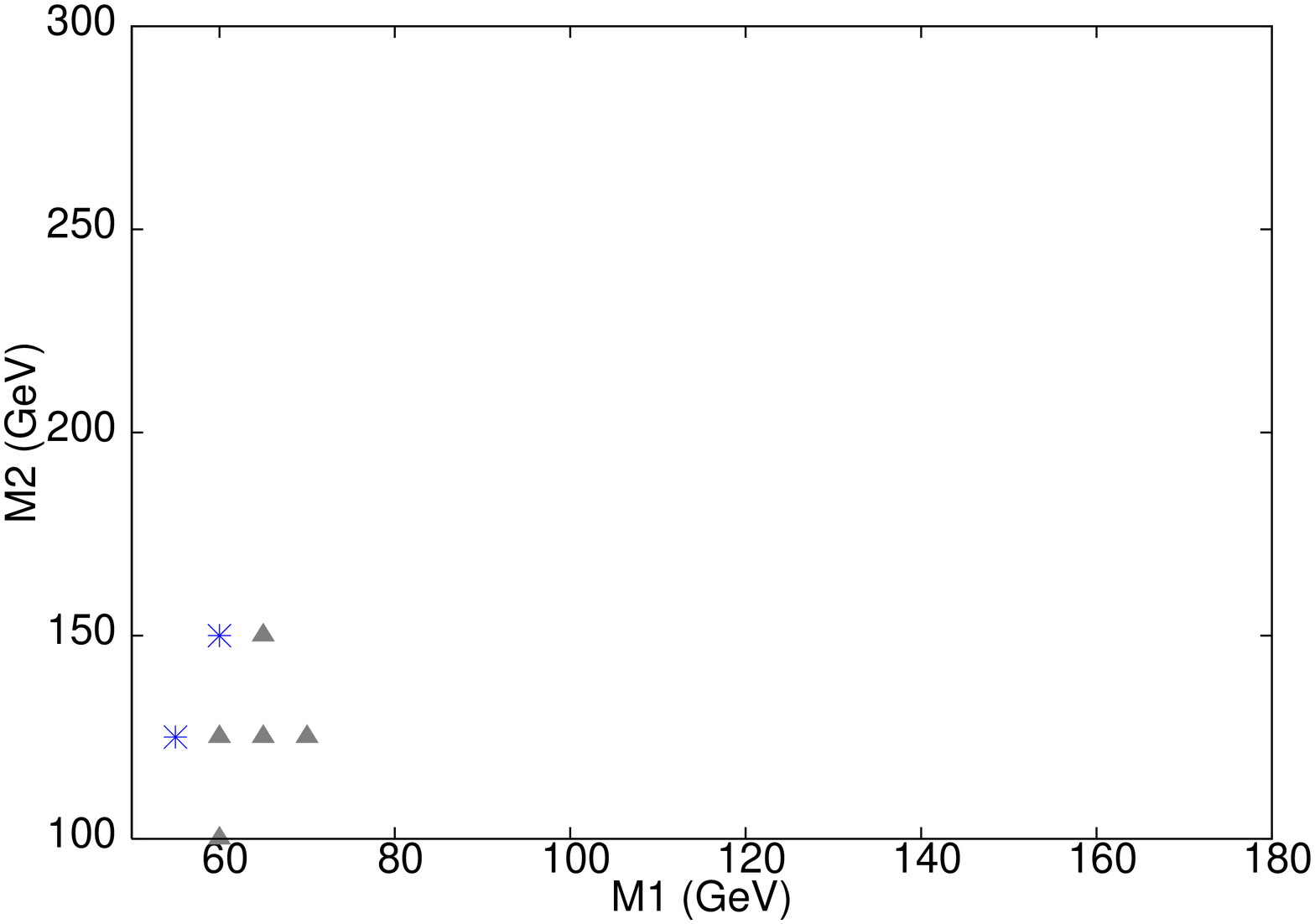,width=6.5 cm,height=7.5cm,angle=-90}}
\vskip 10pt
\centerline{\psfig{file=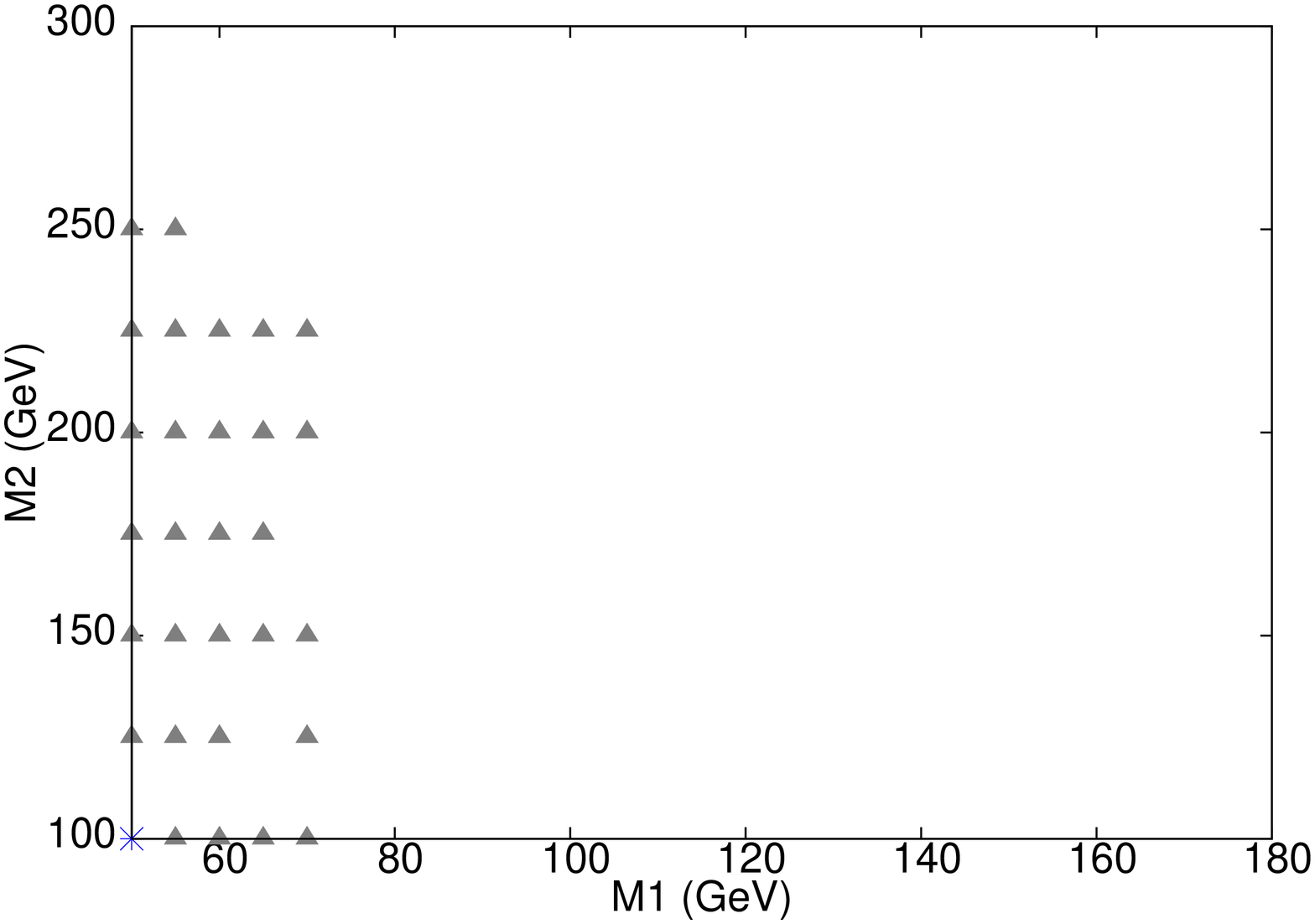,width=6.5 cm,height=7.5cm,angle=-90}
\hskip 20pt \psfig{file=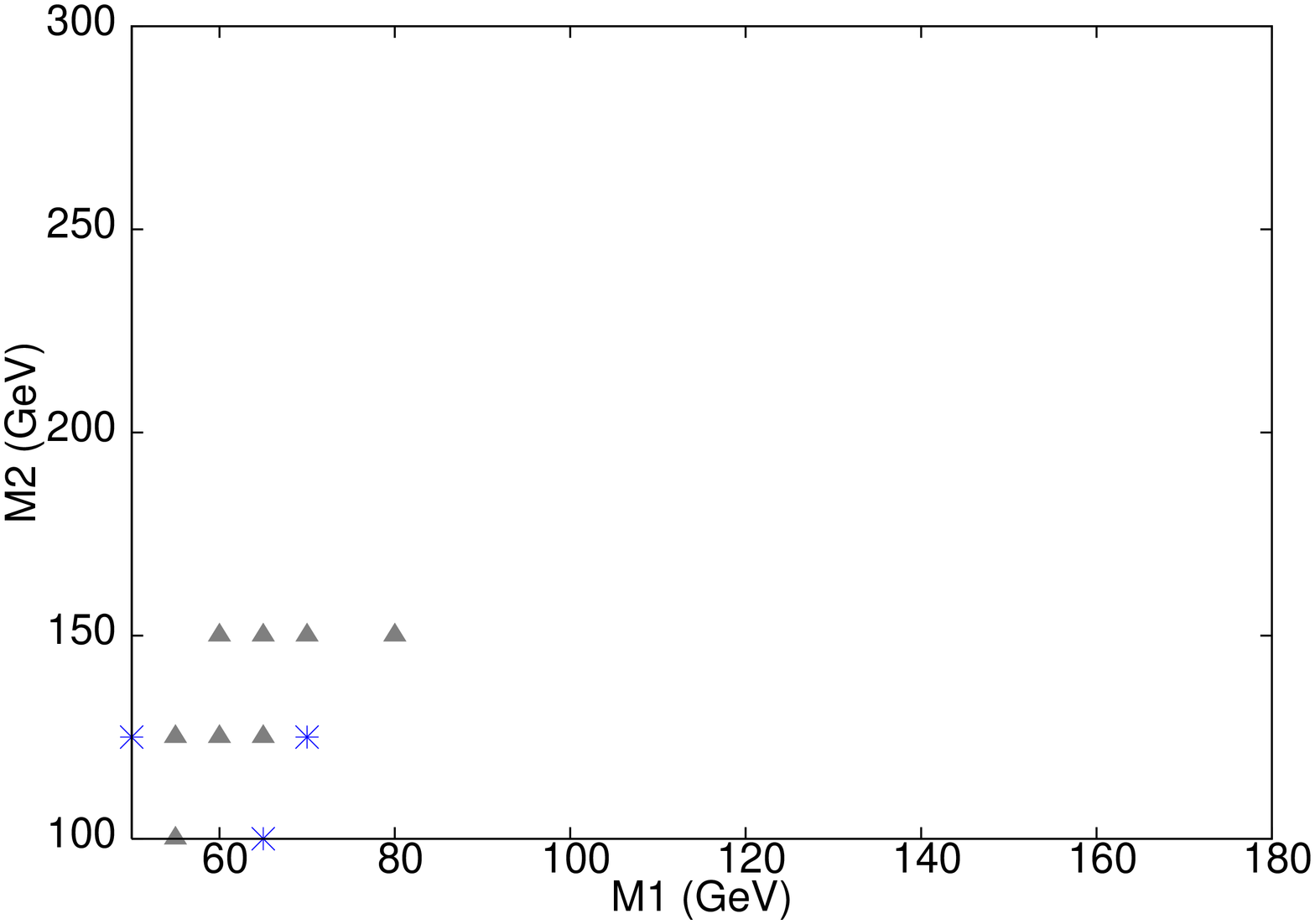,width=6.5 cm,height=7.5cm,angle=-90}}
\caption{Scattered plot of the significance of single-lepton events (on the 
left side) and dilepton events (on the right side) for an integrated 
luminosity of 100 fb$^{-1}$ in $M_1-M_2$ plane. 
{\it Significance Code: {\tt Triangular points:} 1$\le\sigma<$1.5, 
{\tt Star marked points:} $\sigma$$\ge$ 1.5}.
Top row: Slepton mass = 200 GeV, Bottom row: Slepton mass = 400 GeV, 
Left Column: Single-lepton events, Right Column: Dilepton events. 
{\tt CTEQ5L} PDFset used with $\mu_{F}=\mu_{R}=~ \sqrt{\hat s}$.} 
\end{center}
\end{figure}

We also show the variation with $\tan \beta$ from 3 to 20 in Fig 3 in the 
same region 
of parameter space with $m_{\tilde g}= m_{\tilde q}$ = 5 TeV where the 
cross-section decreases sharply. Beyond 20 one 
ends up with a stau LSP, which turns into tachyonic stau state
as  $\tan \beta$ grows larger. The signal rate goes down for higher 
$\tan\beta$. This is because the lighter stau
eigenstate becomes gradually lighter with respect to the other
sleptons, and the decays of the lighter chargino and the second 
lightest neutralino take place more into 
the tau-channels. Again, smaller values of
$\mu$ will affect the signal adversely, since the lighter
chargino/neutralino eigenstates then have enhanced Higgsino components. 
This either tends to open their decays into a Higgs, or causes them to 
decay into final states involving $\tau$'s.

We have discussed above the viability of the hadronically quiet 
trilepton signals at the LHC in terms of statistical significance 
in specific situations. It should
be noted that we have left out the effects of systematic errors here.
When the signal is a few percent of the background, one may have
problems due to systematic shift in the background, especially
if the background is large \cite{Ball:2007zza}. How well the signals can fare 
under such circumstances depends on whether the systematics affect 
the signal and the background strengths in a similar way or not.
In addition, the ultimate success of probes in such a final state 
will depend on the accurate estimate of backgrounds, possibly
in the light of initial data available at the LHC. Since 
this is an open issue, which is serious in much wider context, we 
would just like to keep the reader aware of the need to be cautious on 
this matter.

\section{Other signals}

It may be worthwhile to check whether our benchmark scenario
has accessibility by other types of signals. Table 1 shows the
advantage of the hadronically quiet trilepton signal. However, a scan
over the parameters is required to establish a general conclusion
on the scenario where the coloured suerparticles are too massive
to have any significant contribution to final states at the LHC. 
With this in view, we have studied signals with
$n\ell~+~\ge2~hard~jets~+~\not{E_T}$ across the $M_1 - M_2$ plane,
with the slepton mass set at 200 and 400 GeV respectively. The various
panels in Figure 4 contain the results of this scan. Each of the
hard jets is required to have $E_T \ge 100$ GeV and $|\eta|\le 2.5$, 
the cuts on leptons and $\not{E_t}$ being the same as in the case of
hadronically quiet trileptons.

The figure shows that the single-and dilepton signals both fail
to achieve  significance higher than $2\sigma$ in the entire region of
relevance, with an integrated luminosity of $100~fb^{-1}$. For the
trilepton channel with associated hard jets, it is even less than 1 and have 
not been presented pictorially. Thus in general the other channels are always 
of less advantage than hadronically quiet trileptons, as was suggested at the 
beginning of the paper. The reason behind this is the low event rate from 
gluino/squark production when both of them are very heavy. Thus we are 
essentially dependent on elecroweak processes, where the demand of at least 
two hard central jets has a negative effect. Without such jets, on the other 
hand, one has rather large backgrounds which could be handled in the case of
hadronically quiet trileptons with the help of an invariant mass cut.

We have also checked the effect of reducing the $p_T$ cut on
the hard jets to $75$ and $50$ GeV in succession. It is found that
the significance increases at best by about a factor of two in the 
favourable situatons. However, the uncertainty in backgrounds
increases considerably in such cases.

\section{Conclusions}

In summary, SUSY scenarios with non-universality in both gaugino and 
scalar masses, can envision regions in the parameter
space where the usual signals from the cascade decays of strongly interacting
superparticles involving hard multi-jets drop below the threshold of 
observability.
We demonstrate that hadronically quiet trileptons can be of significant help 
in these cases. As a numerical study presented here indicates,
other signals such as single-or dileptons, for which additional
hard jets are required for background supression, are decidedly
less advantageous for such a scenario. 
Most favourable in this respect are regions
with slepton masses not too far above 200 GeV, and either both
$M_1$ and $M_2$ in the 100 - 200 GeV range, with relatively large production
rates, or with a large separation between them so as to enable the decay- 
leptons to be harder. These two effects yield a substantial
region in the parameter space with 5$\sigma$ or better statistics,
while a still larger region with 3-5$\sigma$ effects can be
identified for an integrated luminosity of 100 fb$^{-1}$. With
higher accumulated luminosity, of course, the reach of the signal
increases. The effects can be expected to be experimentally favourable 
for $\tan\beta \lsim$15 - 20, and with
gaugino-dominated low-lying neutralino and chargino states.\\

\noindent
{\bf Acknowledgment:} 
We thank Abhijit Bandyopadhyay, Paramita Dey, and Satyanarayan Mukhopadhyay 
for technical help. This work was partially supported by funding available 
from the Department of Atomic Energy, Government of India for the 
Regional Centre for Accelerator-based Particle Physics, 
Harish-Chandra Research Institute. Computational work for this study was
partially carried out at the cluster computing facility of
Harish-Chandra Research Institute ({\tt http:/$\!$/cluster.mri.ernet.in}).

\end{document}